\documentclass[12pt]{article}
\usepackage[cp1251]{inputenc}
\usepackage[russian]{babel}
\usepackage{graphicx}

\large \baselineskip = 20 true pt

\begin{document}
\begin{center}
{\Large \bf Realization logical operation on base of the quantum neuron}
\vspace{0.5cm}

{\bf S.V.Belim$^1$, S.Yu.Belim\\
Omsk State University, Russia\\
$^1$belim@univer.omsk.su}
\end{center}

As is well known, learning the formal neuron for the reason
realization logical operation faces with beside difficulties.
These problems are bound, first of all, with problem of the branch.
Some algorithms does not manage to realize on apart taken neuron,
for instance logical scheme "excluding or" (XOR).

Some of the problems leaves if realize the neuron as quantum system.
As any quantum object, neuron is found in vague condition before process of
the measurement. As such process of excitement, translating neuron in determined
condition, possible consider presence of the input signal on synapse of the neuron.
Transition in concrete condition brings about appearance of the certain signal
on leaving the neuron. Before presenting the input signal can be determined only
probability of the finding of the neuron in a certain condition.

We shall Consider two possible realizations of the quantum neuron.
Let condition of neuron is described function of the condition:
\begin{equation}\label{1}
    \Psi=\sum_{i=1}^N c_i\Psi_i,
\end{equation}
here N-amount synapse, $c_i$ -- weighting factors, $\Psi_i$ -- set
orthogonality function:
\begin{equation}\label{2}
    \int \Psi_i^*\Psi_jdV=\delta_{ij},
\end{equation}
 $\delta_{ij}$ -- Kronekers symbol,  $\Psi_i^*$ -- function complex associate of $\Psi_i^*$.

Let $x_i$ -- signal given on  synapse $i$. Input signal we can consider
as condition with function of the condition:
 \begin{equation}\label{3}
    \Phi=\sum_{i=1}^N x_i\Psi_i^*
\end{equation}
Then excitement of the neuron will be defined projection to functions
of the condition of the neuron on condition of the input signal:
\begin{equation}\label{4}
\langle \Phi | \Psi\rangle=\sum_{i,k=1}^Nc_ix_k\int\Psi_i^*\Psi_kdV
\end{equation}
From orthogonality of function we get
\begin{equation}\label{5}
\langle \Phi | \Psi\rangle=\sum_{i,k=1}^Nc_ix_i
\end{equation}
Let  $y$ -- the output signal, taking importance zero if neuron in
nonexcited condition, and unit if in agitated.
We shall Choose as functions of the response step-like
$\Theta$-function:
\begin{equation}\label{6}
    \Theta(\nu)=\left\{%
\begin{array}{ll}
    1, & \hbox{$\nu > 0$;} \\
    0, & \hbox{$\nu \leq 0$.} \\
\end{array}%
\right.
\end{equation}
\begin{equation}\label{7}
    y=\Theta(\langle\Phi|\Psi\rangle-u_0)
\end{equation}
here  $u_0$ -- threshold of excitement. Neuron does not on signals smaller than
$u_0$, and have reaction on signals greater than $u_0$.

Substitution (\ref{5}) in (\ref{7}) gives:
\begin{equation}\label{8}
   y=\Theta(\sum_{i=1}^N c_ix_i-u_0)
\end{equation}
Given model complies with classical neuron, in which occurs the
linear summation input signal with weight.
Тне problem of the education of such neuron is reduced to finding weighting
 factor $c_i$ and threshold of excitement $u_0$. However exists any
problems not allowed within the framework of such neuron.

Other possible realization of the quantum neuron is consideration
of importances of the input signal as quantum чисел of the input condition,
on which is designed neuron. This brings about restriction of possible
importances of the input signal by rational number.

Neuron is described wave function:
\begin{equation}\label{9}
    \Psi=\sum_{n1,...,nk}c_{n1,...,nk}\Psi(n1,...,nk)
\end{equation}
Input signal shall match function of the condition:
\begin{equation}\label{10}
    \Phi=\Psi(n1,...,nk)
\end{equation}
Output signal is formed by means of functions of the response:
\begin{equation}\label{11}
    y=\Theta(\langle\Phi|\Psi\rangle-u_0)
\end{equation}
Education of such neuron is also reduced to choice of importances factor
$c_{n1,...,nk}$ and $u_0$.

We shall Consider the realization to logical operation "excluding OR" (XOR),
impossible within the framework of classical neuron.
Let neuron has two синаптических of the entry,
that is to say input vector  $X=(x_1,x_2)$, $x_i=0,1$.
We shall require that output signal was zero ($y=0$),
if $X=(0,0)$ or $X=(1,1)$, and single ($y=1$), if $X=(1,0)$ or $X=(0,1)$.

As base shall choose the own functions of the system two particles with spin
$1/2$ each. The projection of spin each particles can take importances
$+1/2$ and $-1/2$.  Function of the condition of the neuron, with provision for
principle of nondifferentiality of particles can be recorded as:
\begin{eqnarray}\label{12}
\Psi&=&0\cdot\Psi_1(1/2)\Psi_2(1/2)+0\cdot\Psi_1(-1/2)\Psi_2(-1/2)\\
&+&1\cdot\Psi_1(1/2)\Psi_2(-1/2)+1\cdot\Psi_1(-1/2)\Psi_2(1/2).\nonumber
\end{eqnarray}
The Input signal shall form as follows:
\begin{equation}\label{13}
 \Phi=\Psi_1(x_1-1/2)\Psi_2(x_2-1/2).
\end{equation}
The Output signal shall form same step-like function by means of all:
\begin{equation}\label{14}
    y=\Theta(\langle\Phi|\Psi\rangle-0.5)
\end{equation}
Using condition of normalization
\begin{equation}\label{15}
    \int \Psi_{i1}^*(s_{i2})\Psi_{j1}(s_{j2})dV=\delta_{i1j1}\delta_{i2j2},
\end{equation}
we get the neuron with sought characteristic.

Thereby quantum neuron of the first type equivalent classical neuron.
Quantum neuron of the second type allows to realize all elementary logical functions.
\end{document}